\documentclass[draft]{revtex4-1}[12pt]
\usepackage{graphicx}
\usepackage{color}
\usepackage{bm}
\usepackage{amsmath}
\usepackage{color}
\usepackage{epsfig}
\usepackage{amssymb,amsmath}
\begin{document}

\newcommand\revtex{REV\TeX}
\title{Wave theories of non-laminar charged particle beams:\\ from quantum to thermal regime}
\author{\textbf{Renato Fedele}$^{1,2}$, \textbf{Fatema Tanjia}$^{1,2}$, \textbf{Dusan Jovanovi\'c}$^{3}$,\\ \textbf{Sergio De Nicola}$^{4,2}$, and \textbf{Concetta Ronsivalle}$^5$}

\address{{\small $^1$ Dipartimento di Scienze Fisiche, Universit\'{a} di Napoli ``Federico II" and INFN, Napoli, Italy\\
$^2$ INFN Sezione di Napoli, Napoli, Italy\\
$^3$ Institute of Physics, University of Belgrade, Belgrade, Serbia\\
$^4$Istituto Nazionale di Ottica - C.N.R., Pozzuoli (NA), Italy\\
$^5$Centro Ricerche ENEA, Frascati, Italy}}

\begin{abstract}
{The standard classical description of non-laminar charge particle beams in paraxial approximation is  extended to the context of two wave theories. The first theory we discuss \cite{Fedele1992a,Fedele2011,Tanjia2011}, is based on a quantum-like model, the so-called Thermal Wave Model (TWM) \cite{Fedele1991} that interprets the paraxial thermal spreading of the beam particles as the analog of the quantum diffraction. The other theory,
is the one based on a recently developed model \cite{Fedele2012a,Fedele2012b}, hereafter called Quantum Wave Model (QWM), that takes into account the individual quantum nature of the single beam particle (uncertainty principle and spin) and provides the collective description of the beam transport in the presence of the quantum paraxial diffraction. QWM can be applied to beams that are sufficiently cold to allow the particles to manifest their individual quantum nature but sufficiently warm to make the overlapping of the single-particle wave functions negligible. Compared to QWM,  TWM can be applied to warmer beam, but nevertheless the description of the beam transport and dynamics can be provided with a formalism which is the analog of the quantum one. In both theories, the propagation of the beam transport in plasmas or in vacuo is provided by fully similar set of nonlinear and nonlocal governing equations, where in the case of TWM the Compton wavelength (fundamental emittance) is replaced by the beam thermal emittance. In both models, the beam transport in the presence of the self-fields (space charge and inductive effects) is governed by a suitable nonlinear nonlocal 2D Schr\"{o}dinger equation that is used to obtain the envelope beam equation in quantum and quantum-like regimes, respectively.  An envelope equation of the Ermakov-Pinney type, that includes the collective effects, is derived for both TWM and QWM regimes. In particular, in thermal regime we recover the well known Sacherer equation whilst, in quantum regime, within Hartree's mean field approximation, we obtain the evolution equation of the single-particle spot size, i.e., single quantum ray spot in the transverse plane (Compton regime). We show that such a quantum evolution equation contains the same information carried out by an evolution equation for the beam spot size (description of the beam as a whole). This is done by defining heuristically the lowest QWM state that is reachable by a system of \textit{overlapping-less} Fermions, associated with temperature values sufficiently low to make the single-particle quantum effects visible at the beam scales (but sufficiently high to make negligible the overlapping of the single-particle wave functions). Finally, we show that this lowest QWM state seems to play the role of \textit{borderline} between the description in fundamental single-particle Compton regime and the collective quantum and thermal regimes at larger scales, ranging from nano- to micro-scales (in the beam size). In particular, on the basis of the beam parameters available from some existing advanced compact accelerating machines, it seems to be compatible to the feasibility to reach the nano-sized beams.}
\end{abstract}

\maketitle

%
%
%
%
%
%

\section{Introduction}\label{Intr}
Since its early development, the study of the charged particle motion in electric and magnetic fields has played a special role in the construction of the analogy between optics and mechanics. In fact, problems of charged particle motion, such as the ones occurring in electrostatic and magnetostatic lenses (dipole, quadrupole and other multipoles), magnetic mirrors, magnetic bottles, mass spectrometers and many others \cite{Sturrock1955}, have been used as classic examples to show the correspondence between geometrical optics and classical mechanics. The phenomenological wealth exhibited in this context has generated a separate discipline known as
\textit{electron optics}. Remarkably, within the context of electron optics, important scientific and technological developments, such as the ones in electron microscopy \cite{Sturrock1955,Zworykin1945,Glaser1952,Klemperer1971,Grivet1972,Hawkes1972} and in particle accelerators \cite{Lawson1988}, respectively, have been registered.\newline Thanks to those developments, electron optics has been expanded mainly in two branches that characterized the analogy between optics and mechanics in two different ways also at the level of the wave description.\newline
One of them concerned the behavior of the single particle or systems of a relatively few particles in contexts in which the extension from the \textit{geometrical} to the \textit{wave-like} description simply coincided with the development of the quantum-mechanical description of electron optics, such as in contexts in which the quantum behavior of the single particle prevails over the collective ones and the concept of temperature does not come into play. Valuable examples are encountered in recent studies of vortex states (angular orbital momentum effects and spin) with integer or fractional topological charge that are associated with the single charged-particle matter wave in the electron microscope \cite{McMorran2011}; or in studies of the transport of relativistic charged-particle beams traveling in a high-energy accelerating machines based on the quantum mechanical behavior of the single-particle \cite{Jagannathan1989,Jagannathan1990,Khan1995}. \newline
The other branch of the electron optics concerned, instead, situations in which the behavior of a beam, constituted by an extremely high number of charged particles, is affected by the electromagnetic interactions that are established within such a system. On the other hand, due to the large number of particles, the effects of the temperature cannot be ignored. Therefore, in general, the behavior of such a system is expected to be collective and affected by the thermal spreading among the particles (thermal regimes).
In this regime, electron optics has received a strong development within both conventional and non-conventional accelerator physics. The main applications have regarded the solution of scientific and technological problems concerning the \textit{generation} (charged-particle beam sources such as electron guns), \textit{transport} (pre-acceleration, injection and extraction, electrostatic or electromagnetic acceleration mechanisms, linear or circular accelerating machines), and \textit{use} (medical diagnostics, lithography, free electron laser, colliders, focusing systems, etc.) of charged particle beams \cite{Lawson1988,Chao1998}.
Hereafter, we refer to this branch to as \textit{electron optics in thermal regime} (EOTR).
A peculiar aspect of the paraxial charged-particle beam propagation, as described by the electron optics in thermal regime, is the mixing of the \textit{electron rays} (charged-particle trajectories). In fact, due to the thermal spreading among the particles, the electron rays in a beam, that is traveling \textit{in vacuo} with a relativistic motion, are slightly deviated with respect to the propagation direction with random slopes whose r.m.s. deviation $\sim v_T/c\ll 1$ ($v_T$ and $c$ being the thermal velocity and the light speed, respectively). Furthermore, the thermal spreading introduces an uncertainty in the electron ray positions in the transverse plane at any longitudinal position. Then, the picture that comes from the envelope of the above electron ray mixing resembles the one that is exhibited by the paraxial ray diffraction in the light rays of an electromagnetic radiation beam.\newline
It is well known that from this comparison one can conclude that the statistical behavior of the charged particles in a paraxial beams, which is a fully classical process, \textit{simulates} the paraxial diffraction exhibited by the electromagnetic radiation beams. Its experimental evidence is manifested in all the processes that are relevant to EOTR \cite{Lawson1988} and is supported by theoretical kinetic descriptions (Boltzmann/Vlasov equation) \cite{Lawson1988}. An alternative  theoretical description has been proposed in Ref. \cite{Fedele1991} by extending the EOTR in paraxial approximation to wave context. Within the context of the analogy between optics and mechanics, a quantization procedure to transit from geometrical to wave description of EOTR have been performed \cite{Fedele1991}. It was the first time that EOTR was formulated in terms of a wave description. This approach was formulated in a way fully similar to the one of Gloge and Marcuse to transit from the geometrical to wave formulation of e.m. radiation optics \cite{Gloge1969}. In turn, the Gloge and Marcuse procedure was formulated in a way fully similar to the one of Bohr to construct the quantum mechanics.
In Refs. \cite{Fedele1991} it was shown that, within the context of the wave description of EOTR, the transverse dynamics of relativistic charged particle beam in paraxial approximation is governed by a 2D Schr\"{o}dinger-like equation for a complex function, called beam wave function (BWF), whose squared modulus is proportional to the particle density profile. Here the propagation direction plays the role of time-like variable and the thermal beam emittance replaces the Planck's constant. This \textit{quantum-like} approach is usually referred to as Thermal Wave Model (TWM) \cite{Fedele1991}. With the use of TWM, a number of linear and nonlinear problems in both conventional and plasma-based particle acceleration were successfully described \cite{Fedele1992a,Fedele1992b,DeNicola1995,Fedele1993,Fedele1994a,Fedele1994b,Fedele1995}. In particular, TWM has been applied to the Gaussian particle-beam optics and dynamics for a quadrupole-like device \cite{Fedele1991}, to luminosity estimates in final focusing stages of linear colliders in the presence of small aberrations \cite{Fedele1992b}, whilst the TWM predictions have been compared with tracking-code simulations and a fair agreement has been demonstrated (the analysis has been carried out in both configuration space and phase space \cite{Fedele1994a,Fedele1995,Jang2007,Jang2010}). Remarkably, a self consistent theory of the interaction between a relativistic electron (positron) beam and a cold unmagnetized plasma has been also developed \cite{Fedele1992a}. Recently, the TWM description of the self-consistent beam-plasma interaction has been also developed in strongly magnetized plasmas, where the collective vortex beam states (orbital angular momentum states) has been predicted \cite{Fedele2011,Tanjia2011}.
The approach was very soon successfully applied to the longitudinal beam dynamics. For instance, it was useful to predict soliton-like states of the charged-particle beams or to provide a wave key of reading for the nonlinear and collective effects exhibited by the coherent instabilities in high-energy accelerating machines, showing that it can be formulated as the deterministic or the statistical approach to modulational instability, where a  Landau-type damping plays a basic role \cite{Fedele1993,Anderson1999,Fedele2000,Johannisson2004}.

More recently, the geometrical electron optics has been extended to the quantum wave context where the collective  interactions among the particles are not neglected \cite{Fedele2012a,Fedele2012b,Jovanovic2012b}. This way, a quantum aprroach, hereafter called Quantum Wave Model (QWM), to relativistic charged particle beams has been proposed.  In this approach, the space charge effects (both capacitive and inductive) are taken into account as in the TMW. In general, the collective effects are taken into account within the Hartree's mean field approximation, as it has been done in TMW. However, the above QWM takes into account the quantum nature of the single particle, such as the single-particle uncertainty principle and the spin of the single particle. However, the collective quantum nature of the system related to the overlapping of the single-particle wave functions is disregarded. So that, for the typical densities of charged particle beams employed in the present generation of both conventional accelerators and plasma-based acceleration schemes,  QWM is appropriate when the temperatures of the beam are sufficiently low to preserve the observability of the individual quantum nature of the particles, but sufficiently high to make the overlapping of the single-particle wave functions negligible (system of overlapping-less wave functions). These physical conditions characterize the paraxial-ray approximation provided by QWM. In fact, if the quantum uncertainty of the single particle is not hidden by the thermal spreading, within a paraxial picture one can attribute to the single electron ray an uncertainty in both position along the transverse plane and slopes with respect to the propagation direction. It results that now the rays are mixed due to the individual quantum nature of the particles and this picture corresponds to the analog of the paraxial diffraction of the light rays in radiation beams. Since here the diffraction that is exhibited is the quantum one, we call \textit{quantum paraxial diffraction} the picture produced by the mixing of the electron rays due to the single-particle uncertainty. \cite{Fedele2012a,Fedele2012b,Jovanovic2012b,Jovanovic2013}. Let us refer hereafter such a beam to as \textit{quantum paraxial beam} (QPB). In this physical situation, the Bohr/Gloge and Marcuse quantization procedures teach us the way to construct the appropriate equation which governs the spatio-temporal evolution of a QPB. It turns out that such a governing equation is a 2D spinorial Schr\"{o}dinger equation. QWM has been recently applied to describe the self-interaction of an electron or positron beam propagating in a strongly magnetized plasma. Quantum ring solitons as the 2D quantum vortex states associated with the angular momentum (orbital plus spin) states have been found as well as the self-focusing conditions predicted \cite{Fedele2012a,Fedele2012b,Jovanovic2012b,Jovanovic2013}.

In this paper, we try to establish a \textit{borderline} condition between the single-particle quantum description (Compton scale) and the collective one in quantum or thermal regimes at larger scales. This is done by considering the charged-particle beam transport, in the quantum and in the thermal regime, respectively, under the action of both the space charge effects and an elastic force acting radially in the transverse plane. We show that in cylindrical symmetry the governing equations for both cases are formally the same, provided that we replace the Compton wavelength with the thermal emittance to transit from the QWM equation to the TWM equation. Then, we approach our problem in a unified way, by starting from a Schr\"{o}dinger-like equation with a generic dispersion parameter (i.e., Compton wavelength or emittance, respectively). In principle, this  Schr\"{o}dinger-like equation is coupled with a Poisson's like equation for the effective potential that accounts for the collective effects.
By using the virial equations we obtain an envelope equation describing the time evolution of the beam spot size (rms of the transverse particle distribution) that is specialized for the quantum and thermal cases, respectively. Then, we consider a nano-sized electron or positron beam as an ensemble of \textit{overlapping-less} single particle wave functions. Since electrons or positrons are Fermions, for a suitable temperature conditions, Pauli's exclusion principle helps us to assume that the beam of such particles constitute a system of overlapping-less  particles. However, to fulfil this assumption, we consider temperatures whose lowest value corresponds to the conditions for which the beam particles are distributed in the $\mu$-space (Boltzmann phase-space) in such a way that, roughly, the quantum spot, corresponding to each of them, touches the nearest ones at their effective border. Since, each effective spot extent is $\sim \lambda_c$, at that lowest temperature the accessible area of the transverse phase space  corresponding to the beam will be roughly $\sim N_\perp \lambda_c$, where $N_\perp$ is the number of spots in the transverse plane (i.e., this number divided by the transverse beam spot size in the real space gives the transverse particle density). For temperatures more and more above this limit (heating of the beam), the single-particle quantum nature no longer appears, being hidden by the thermal spreading, then the transverse phase space spot corresponding to the beam as a whole has an effective extent corresponding to the thermal emittance, say $\epsilon$. Consequently, for Fermions, the lowest value of the beam emittance roughly corresponds to $N_\perp\lambda_c$.
We will show that for number of particles, density and temperature suitably consistent with the existing compact accelerators, we obtain beam spot size in the ranges of nonometers (those that are called \textit{nanobeams}). We finally show that the envelope equation, including the space charge effects, that is obtained from QWM (\textit{Compton scale envelope equation}) is equivalent to the one that holds for the nanoscales (\textit{nano scale envelope equation}), which is obtained just rescaling the beam spot size and the emittance. Using similar arguments, we show the compatibility of those envelope equations with the one at the thermal regime (\textit{thermal scale envelope equation}), usually referred to as Sacherer's equation.
\section{Main equations for the quantum electron/positron beam in the strongly nonlocal regime \label{MainEqns} }
In order to describe our problem with the inclusion of the collective effects in the strong nonlocal regime, let us consider two cases of beam propagation. One is the propagation through a collisionless, cold plasma where the beam experiences the effects of the wake fields that itself produces. The other one is the propagation \textit{in vacuo} where the beam experiences the effects of the space charge (inductive as well as capacitive) that itself produces. In both cases, we assume that the beam is cylindrically symmetric and is travelling along the $z$ axis with the speed $\beta c$. We additionally assume that the beam number density is $\rho_b (r,z,t)=\rho_b (z,\xi)$, where $r$ is the radial cylindrical coordinate and $\xi=z - \beta c$ is the self-similar variable which plays the role of a time-like variable. Due to this assumption, we also assume that all the quantities generated by the beam (such as the density current and the e.m. fields in the plasma and/or in vacuo) have the same kind of spatio-temporal dependence.  The connection between the beam density, i.e., $\rho_b (r,\xi)$, and the BWF has to be established for both TWM and QWM.

\subsection{Propagation through a plasma\label{PPL}}
Referring to our previous papers \cite{Fedele1992a,Fedele2011,Tanjia2011,Fedele2012a,Fedele2012b,Jovanovic2012b,Jovanovic2013}, the transverse nonlinear and collective quantum-like (TWM) or quantum (QWM) dynamics of a cylindrically symmetric relativistic electron/positron beam, propagating at speed $\beta c$ ($\beta\simeq 1$) in a plasma in the overdence conditions (beam density much smaller than the plasma densitity), is governed by the following pair of equations:
\begin{eqnarray} &&i\alpha\frac{\partial \psi_m}{\partial\xi}=
-\frac{\alpha^2}{2}\frac{1}{r}\frac{\partial}{\partial
r}\left(r\frac{\partial \psi_m}{\partial
r}\right)+U_w\psi_m
+\left(\frac{1}{2}Kr^2+\frac{m^2\alpha^2}{2r^2}\right)\psi_m,
\label{c3}\\
&&\frac{1}{r}\frac{\partial}{\partial
r}\left(r\frac{\partial U_w}{\partial
r}\right)-\frac{k^4_{pe}}{k_{uh}^2}\,U_w=
\frac{k^4_{pe}}{k_{uh}^2}
\frac{\rho_b}{n_0\gamma_0}\,, \label{c4}\
\end{eqnarray}
where the plasma is supposed strongly magnetized by the action of an external uniform and static magnetic field $\mathbf{B}_0$ along the propagation direction of the beam, the ions are supposed to be infinitely massive, $k_{pe}  =\omega_{pe}/c$ is the the plasma wavenumber ($\omega_{pe}$ being the electron plasma frequency),  $k_{uh}=\omega_{uh}/c$, ($\omega_{uh}$ being the upper hybrid frequency), $n_0$ is the unperturbed plasma number density, and $m_0\gamma_0c^2$ is the unperturbed longitudinal relativistic total energy of the single particle of the beam. Here, $m$ is an integer related to the eigenvalues of the orbital angular momentum (\textit{vortex charge}), $U_w (r,\xi)$ is the plasma wake potential as a results of the collective beam-plasma interaction, and $K=\left(eB_0/2m_0\gamma_0c^2\right)^2$ is the elastic constant introduced by the external magnetic field that produces a quadratic trapping potential well for the beam particles along the transverse plane. Equation (\ref{c4}) has been obtained in the long beam limit of the plasma wake field excitation, i.e., $k_{pe}\sigma_z \gg 1$ ($\sigma_z$ being the beam length).\newline
Note that $\alpha$ is a generic dispersion parameter: if it is equal to the thermal emittance, i.e., $\alpha = \epsilon$, then it describes the beam dynamics within TWM \cite{Fedele1991}; otherwise, if we assume that it equals the relativistic Compton wavelength, i.e., $\alpha = \lambda_c/\gamma_0 \equiv \epsilon_c$, then it describes the beam dynamics within QWM \cite{Fedele2012a,Fedele2012b}. A relationship between the BWF $\psi_m (r,\xi)$ and $\rho_b(r,\xi)$ has to be established in connection with the particular regime (thermal or quantum). In fact, when $\alpha=\epsilon$ and $\psi_m(r,\xi)$ is suitably normalized, at any $\xi$, $|\psi_m (r,\xi)|^2$ represents the probability density of electron rays in the transverse plane. However, the uncertainty in the transverse electron ray position is attributed to the paraxial thermal spreading in TWM, and $\psi_m (r,\xi)$ describes the spatio-temporal evolution of the beam as a whole. When $\alpha=\epsilon_c$, $\psi_m (r,\xi)$ represents the single-particle wave function. Then, $|\psi_m (r,\xi)|^2$ does not describe the beam as a whole. However, the Hartee's mean field approximation fixes the connection with the ``macroscopic'' scales and, therefore, with the density. Then in QWM, each electron ray is affected by the quantum uncertainty in the ray position on the transverse plane as well as in each slope with respect to the propagation direction.
Note that Eqs. (\ref{c3}) and (\ref{c4}) describes the self-consistent interaction of the beam with the plasma. The beam density, through the term at the right hand side of Eq. (\ref{c4}), excites the plasma wake which, in turn, acts on the beam propagation through the potential term of Eq. (\ref{c3}).

We confine our attention on the case of zero vortex charge ($m=0$) and in the \textit{strong nonlocal regime} \cite{Fedele2012a,Jovanovic2012b}, i.e.,
$$\frac{1}{r}\frac{\partial}{\partial
r}\left(r\frac{\partial U_w}{\partial
r}\right) \gg \frac{\omega^2_{pe}}{\omega^2_{uh}}
\frac{\omega^2_{pe}}{c^2}\,U_w\,,$$
then, the pair of equations (\ref{c3}) and (\ref{c4}) reduces to
\begin{eqnarray}
&&i\alpha\frac{\partial \psi}{\partial\xi}=
-\frac{\alpha^2}{2}\frac{1}{r}\frac{\partial}{\partial
r}\left(r\frac{\partial \psi}{\partial
r}\right)+U_w\psi
+\frac{1}{2}Kr^2\,\psi,
\label{c3-1}\\
&&\frac{1}{r}\frac{\partial}{\partial
r}\left(r\frac{\partial U_w}{\partial
r}\right)= \mu_p\,\rho_b\,,\label{c4-1}\
\end{eqnarray}
where we have used the substitution $\psi_0 \rightarrow \psi$ and $\mu_p=k^4_{pe}/{k_{uh}^2}n_0\gamma_0$. The second one can be easily integrated as
\begin{equation}\label{Poisson2}
U_w=\mu_p\int_0^{r}\frac{1}{r^\prime}\,dr_\perp^\prime \int_0^{r^\prime}\rho_b(r^{\prime\prime},\xi)\,r^{\prime\prime}dr^{\prime\prime}\,dr'\,,
\end{equation}
where the arbitrary integration constant that comes out from the first integration has been put equal to zero to impose the non-divergence of $U$ in $r=0$, whilst the additive arbitrary constant appearing from the second integration has been fixed to zero without loss of generality.

\subsection{Propagation in vacuo\label{PVC}}
If the beam is travelling in vacuo, then each single particle experience a total Lorentz force constituted by both an electric and a magnetic parts. In fact, the beam space charge generates an electric field within the beam itself that, in the cylindrical symmetry we have assumed, is radial. Moreover, if the beam is traveling, then the space charge generates a current density along the propagation direction, say $z$, which is the source of a magnetic field within the beam itself; due to the cylindrical symmetry this magnetic field is azimuthal. It is easy to see that the electric and magnetic forces on each particle are repulsive and attractive, respectively, along the radial direction, so that the total Lorentz force experienced by each beam particle comes from the interplay of these two forces. In particular, the electric force is compensated by the magnetic one to the order of $1/\gamma_0^2$. This implies that the space charge blow up reduces as the beam energy increases. It is easy to see that, by using the Maxwell's equations and introducing the self-similar variable $\xi$, the propagation of the beam in vacuo but in the presence of the external longitudinal magnetic field $\mathbf{B}_0$, is governed by the following pair of equations
\begin{eqnarray}
&&i\alpha\frac{\partial \psi}{\partial\xi}=-\frac{\alpha^2}{2}\frac{1}{r}\frac{\partial}{\partial
r}\left(r\frac{\partial \psi}{\partial r}\right)+U_{sp}\psi+\frac{1}{2}Kr^2\,\psi,\label{c3-2}\\
&&\frac{1}{r}\frac{\partial}{\partial r}\left(r\frac{\partial U_{sc}}{\partial r}\right)= \mu_{sc}\,\rho_b\,,\label{c4-2}\
\end{eqnarray}
where $U_{sc}(r,\xi)$ stands for the effective space charge potential (normalized with respect to $m_0\gamma_0c^2$ experienced by the beam particles and $\mu_{sc}=-2\pi e^2/m_0\gamma_0^3 c^2$. Then, Eq. (\ref{c4-2}) can be readily integrated as Eq. (\ref{c4-1}), viz.
\begin{equation}\label{Poisson-sc}
U_{sc}=\mu_{sc}\int_0^{r}\frac{1}{r^\prime}\,dr_\perp^\prime\int_0^{r^\prime}\rho_b(r^{\prime\prime},\xi)\,
r^{\prime\prime}dr^{\prime\prime}\,.
\end{equation}

\subsection{Unified description\label{UD}}
To describe in a unified way both the above cases presented in sections \ref{PPL} and \ref{PVC} in TWM and QWM, respectively, in the strong nonlocal regime, let us start by the pair of equations
\begin{eqnarray}
&&i\alpha\frac{\partial \psi}{\partial \xi}=
-\frac{\alpha^2}{2}\frac{1}{r}\frac{\partial}{\partial
r}\left(r\frac{\partial \psi}{\partial
r}\right)+V(r,\xi)\psi,
\label{c3-3}\
\end{eqnarray}
where
\begin{eqnarray}
&&V(r,\xi)=U(r,\xi)+\frac{1}{2}Kr^2\,,\label{v1}\\
&&U(r,\xi) =\mu\,\int_0^{r}\frac{1}{r^\prime}\,dr^\prime\int_0^{r^\prime}\rho_b(r^{\prime\prime},\xi)\, r^{\prime\prime}dr^{\prime\prime}\,.\label{b2}\
\end{eqnarray}
This way, we can consider different possible combinations of $\alpha (= \epsilon, \epsilon_c)$ and $\mu (= \mu_p, \mu_{sc})$.

\section{Aberrationless solutions\label{AS}}
According to the strong non local regime, the beam can experience, in principle, strong focusing effects \cite{Jovanovic2012b}. Then, we can assume that its density $\rho_b(r,\xi)$ is sufficiently peaked around the propagation direction (i.e., $r=0$). So that, the potential well $V(r,\xi)$ can be suitably expanded in $r$ around the propagation direction up to the second power. This way, the functional dependence of the BWF, solution of Eq. (\ref{c3-3}), will be determined according to the expansion of $V(r,\xi)$. Note that, from Eq. (\ref{b2}), $U(0,\xi)=0$ and therefore $V(0,\xi)=0$, as well. In addition, since we have assumed that $\rho_b(r,\xi)$ has a maximum in $r=0$, we conveniently assume that $V$ has a minimum in $r=0$, namely $\left(\partial V/\partial r\right)_{r=0} =0$, and $\left(\partial^2 V/\partial r^2\right)_{r=0} > 0$. It is easily seen that, once we impose the first of these conditions, the secon-order derivative of $V$ in $r=0$, becomes, viz.,
%
%
$$\overline{K}(\xi)\equiv\left(\frac{\partial^2V}{\partial r^2}\right)_{r=0}=K+\frac{\mu}{2}\rho_{max}\,.$$ where $\rho_{max}=\rho_b(r=0,\xi)$ is the maximum density value. Note that we can estimate this maximum value as $\rho_{max}=N/\pi\sigma_z\sigma^2$, where $N$ is the total number of the beam particles and $\sigma$ is the (transverse) beam spot size. Therefore,
\begin{equation}\label{Kbar}
\overline{K}(\xi)=K+\frac{\mu N}{2\pi \sigma_z\sigma^2(\xi)}.
\end{equation}
Note that $\overline{K}$ is a function of $\xi$ because $\sigma$ depends on $\xi$ during the beam evolution. Therefore, to get the aberrationless solutions we have to solve the following Schr\"{o}dinger-like equation.
\begin{equation}
i\alpha\frac{\partial \psi}{\partial\xi}=-\frac{\alpha^2}{2}\frac{1}{r}\frac{\partial}{\partial r}\left(r\frac{\partial \psi}{\partial r}\right)+\frac{1}{2}\overline{K}(\xi)r^2\,\psi\,.\label{g1}\
\end{equation}
The positivity of $\overline{K}(\xi)$ is also required to have a bounded states. We note that
\begin{equation}\label{K-bar-positivity}
\overline{K}(\xi)>0\,,\,\mbox{for}\,\,\, \mu=\mu_p\,\,\, \mbox{or for}\,\,\, \mu_{sc}>-2K/\rho_b(r=0,\xi)\,.
\end{equation}
Provided that conditions (\ref{K-bar-positivity}) are satisfied, a complete set of normalized solutions of Eq. (\ref{g1}) is given by:
\begin{eqnarray}
\psi_{n}^{(\alpha)}(r,\xi)=\frac{1}{\sqrt{\pi}\sigma_\alpha}
\,\exp\left(-\frac{r^2}
{2\sigma_\alpha^2}+\frac{ir^2}{2\alpha\rho}+i\phi_{n}^{(\alpha)}\right)
L_n\left(\frac{r^2}
{\sigma_\alpha^2}\right)\,, \label{g8}\
\end{eqnarray}
where $n$ integer, $L_n$ are the simple Laguerre polynomials, and $\sigma_\alpha =\sigma_\alpha
(\xi)$, $\rho_\alpha =\rho_\alpha (\xi)$ and $\phi_{n}^{(\alpha)} =\phi_{n}^{(\alpha)} (\xi)$
satisfy the following differential equations
\begin{eqnarray}
&&\frac{1}{\rho_\alpha}=\frac{1}{\sigma_\alpha}\frac{d\sigma_\alpha}{d\xi}\,, \label{g3}\\
&&-\frac{\sigma_\alpha^2}{2\alpha}\frac{d\phi_{n}^{(\alpha)}}{d\xi} = n+\frac{1}{2}\,, \label{g4}\\
&&\frac{d^2\sigma_\alpha}{d\xi^2}+\overline{K}\sigma_\alpha-\frac{\alpha^2}{\sigma_\alpha^3}=0\,.
\label{g5}\
\end{eqnarray}
Note that $\sigma_\alpha(\xi)$ could be defined as the rms position associated with the fundamental mode $\psi_0^{(\alpha)}(r,\xi)$, viz.,
$$
\sigma_\alpha(\xi) = \langle r^2 \rangle^{1/2} \equiv \left[2\pi\int_{0}^{\infty}r^2 |\psi_0^{(\alpha)}(r,\xi)|^2\,r\,dr\right]^{1/2}\,.
$$
Then: for $\alpha=\epsilon$ (TWM), $\sigma_\alpha$ represents the beam spot size ($\sigma_\epsilon = \sigma$); for $\alpha =\epsilon_c$ (QWM), it corresponds to the effective physical extent of the single quantum particle (i.e., single quantum ray) in the real transverse space. By substituting the expression of $\overline{K}$ given by Eq. (\ref{Kbar}) into Eq. (\ref{g5}), we readily get
\begin{equation}\label{Sacherer-eq}
\frac{d^2\sigma_\alpha}{d\xi^2}+K\sigma_\alpha +\eta\,\frac{\sigma_\alpha}{\sigma^2}-\frac{\alpha^2}{\sigma_\alpha^3}=0\,,
\end{equation}
where $\eta= \mu N/2\pi\sigma_z$.\newline
Let us now discuss the role that Eq. (\ref{g5}) plays in both thermal and quantum regimes.

\subsection{Thermal paraxial regime}
According to TWM, the behavior of a paraxial beam is the analog of a single quantum particle. In fact, the beam, as a whole, is described in terms of a wave function that is solution of the Schr\"{o}dinger-like equation (\ref{g1}) with $\alpha =\epsilon$. During its evolution, in each subspace of the transverse phase space, i.e., ($x,p_x$) and ($y,p_y$), the beam as a whole is represented by a spot whose rms extent is $\sim \epsilon$. This is also the same uncertainty that relates position and slope of each electron ray caused by the thermal spreading. In the previous section we have pointed out that $\sigma$ appearing in Eqs. (\ref{g8})-(\ref{g5}) represents the transverse beam spot size, i.e., the rms associated with the fundamental mode $\psi_0$.  Then, since $\sigma_\epsilon=\sigma$, the envelope equation (\ref{Sacherer-eq}) reduces to the following Ermakov-Pinney type equation (\ref{g5})
\begin{equation}\label{Sacherer-eq-1}
\frac{d^2\sigma}{d\xi^2}+K\sigma + \frac{\eta}{\sigma}-\frac{\epsilon^2}{\sigma^3}=0\,.
\end{equation}
The latter exactly recovers the beam envelope equation in the presence of a transverse focusing force and the collective effects, usually referred to as the Sacherer's equation \cite{Lawson1988,Chao1993}, well known  also in the conventional particle beam physics.

\subsection{Quantum paraxial regime}
By definition, the existence of collective effects means the presence of many particles in the beam. Then, according to QWM, a quantum paraxial beam in phase space is characterized by a system of single-particle quantum spots. In the ($x,p_x$) and ($y,p_y$) subspaces of the transverse phase space, each quantum particle (or, equivalently, each quantum electron ray) is represented by a spot whose extent is of the order of $\lambda_c$ (Compton scale).  Let us denote by $\sigma_c$ the transverse rms of the quantum single-particle wave function, which is solution of Eq. (\ref{g1}). Within the Hartree's mean field approximation \cite{Hartree1957}, each quantum particle experiences the collective effects. However, according to the previous results, the collective force in the linear approximation is proportional to $1/\sigma^2$, where $\sigma$ is the transverse beam spot size which is much greater than $\sigma_c$. Therefore, envelope equation (\ref{Sacherer-eq}) in this case becomes
\begin{equation}\label{Sacherer-eq-2}
\frac{d^2\sigma_c}{d\xi^2}+K\sigma_c +\eta\,\frac{\sigma_c}{\sigma^2}-\frac{\epsilon_c^2}{\sigma_c^3}=0\,,
\end{equation}
Let us consider now another important aspect of our charged particles. We have assumed actually that they are electrons or positrons. Therefore, they are Fermions and obey to Pauli's exclusion principle. On the other hand, by construction, QWM assumes that the overlapping of the single-particle wave functions is negligible. It follows that keeping the system sufficiently diluted, sufficiently high temperature values allow the system to keep, in the transverse phase space, the single quantum spots sufficiently  far each other to cover a total phase space area, whose extent is the transverse beam emittance, i.e. $\epsilon$, at least greater than the single quantum spot extent: $\epsilon > \epsilon_c$.

\section{The borderline between single-particle and collective regimes}
Referring to the above arguments, temperature values higher and higher will hide the single-particle quantum effects (uncertainty of the quantum electron rays and spin) more and more ($\epsilon \gg \epsilon_c$). This way the beam transverse dynamics is dominated by the thermal regime where TWM holds. If, vice versa, we reduce the transverse temperature, the beam becomes \textit{sufficiently cold} to reduce the mean distance among the phase space spots to reach the condition, consistent with the exclusion principle as well, at which each of them is roughly confining with the nearest neighbors with their effective borders. In such a condition, the single-particle wave functions are not yet effectively overlapped. Then, the corresponding temperature is roughly the threshold below which QWM is no longer applicable. Then, one has to extend it with the inclusion of the overlapping of the wave functions for a system of fermions. Let us, for the time being, consider the system at this lowest QWM state. Then, it is easy to see that, if $N_\perp$ is the number of quantum electron rays that are intersecting the transverse plane (i.e., number of particles distributed in a single transverse plane), this is also the number of quantum spots belonging to the accessible region in the transverse phase space at any temperature. In particular, this is also the number at that lowest QWM state. Consequently, the phase space area of the accessible region corresponding to the lowest QWM state is roughly $N_\perp$ times the single fundamental emittance, i.e.,
\begin{equation}\label{lowest-state1}
\epsilon^*\approx N_\perp\epsilon_c.
\end{equation}
According to QWM, the envelope equation (\ref{Sacherer-eq-2}) is applicable also to this state. However, by using arguments similar to the ones that have led to condition (\ref{lowest-state1}), we can easily see that, roughly
\begin{equation}\label{lowest-state2}
\sigma^*\approx\sqrt{N}_\perp\sigma_c.
\end{equation}
Then, by using this estimate into Eq. (\ref{Sacherer-eq-2}), we obtain
\begin{equation}\label{lowest-state3}
\frac{d^2\sigma^*}{d\xi^2}+K\sigma^* +\,\frac{\eta}{\sigma^*}-\frac{(N_\perp\epsilon_c)^2}{\sigma^{*3}}=0\,,
\end{equation}
which shows that Eq. (\ref{Sacherer-eq-2}), although written for the quantum single-particle averaged motion in the presence of collective effects, is fully equivalent to a macroscopic equation written for the beam spot size at the lowest state. Note that the presence of $N_\perp$ in the last term of the left-hand side represents a sort of \textit{amplification} of the quantum nature of the beam particles that is related to the Compton scale. This means that it provides a macroscopic manifestation of the single-particle quantum effects. According to Eq. (\ref{c3-1}), the conservation of the BWF norm implies that $N_\perp$ is a conserved quantity.

Note that the ratio $N_\perp/\pi\sigma^{*2}$ estimates the 2D particle number density in the transverse plane. Then, for a beam sufficiently cold such that those quantum effects are visible, we have to require that $\epsilon_N^* \approx N_\perp\lambda_c$, where $\epsilon_N^*=\gamma_0\epsilon^*$ is the normalized emittance at the lowest QWM state. Therefore, we have to choose $\epsilon_N$ in the range:
\begin{equation}\label{emittance-condition-1}
\sqrt{2}\,\lambda_c\,\sigma^*n_b^{1/3}\ll\epsilon_N^*\approx N_\perp\lambda_c\,,
\end{equation}
which implies the condition
\begin{equation}\label{N_perp-condition}
\frac{\sqrt{2}\,\sigma^*n_b^{1/3}}{N_\perp}\ll 1\,,
\end{equation}
where $n_b$ is the unperturbed beam density. The inequality in (\ref{emittance-condition-1}) comes from the condition for which the single-particle wave functions are in the \textit{overlapping-less} condition (i.e., the mean inter-particle distance $\delta \sim n_b^{-1/3}$ is much greater than the thermal de Broglie wavelength $\lambda_{T}=h/m_0\gamma_0 v_{T}$, $v_T$ being the particle thermal velocity).
\section{Feasibility of the lowest QWM state: Nanobeams}
In this section, we want to show the feasibility of the lowest QWM state through some estimates of the beam parameters presently involved in advanced accelerating machines, such as the compact linear colliders (CLIC). 
Starting from the choice of very low emittance and spot size, we assume that our beam is a high-energy nanobeam, i.e., $\gamma_0\sim 10^{5} - 10^{6}$, $\sigma^* \sim 5\times 10^{-9} \mbox{m}$ and we choose $\epsilon_N^* \sim 3\times 10^{-9} \mbox{m rad}$. We assume also that the beam is \textit{long}, i.e., $\sigma_z \gg \sigma^*$. For instance we may assume that $\sigma_z \sim 10^{-4} - 10^{-3}\mbox{m}$. By using the above relationships, we easily relate the de Broglie wavelength to the Compton wavelength, viz.,
\begin{equation}\label{deBroglie}
\lambda_T \approx \sqrt{2}\lambda_c \frac{\sigma^*}{\epsilon_N^*}\,.
\end{equation}
Therefore, since we have chosen $\sigma^* \sim \epsilon_N^*$, we have $\lambda_T \sim \lambda_c$. Consequently, from the relationship at the right hand side of condition (\ref{emittance-condition-1}) we get $N_\bot \sim 1.2\times10^{3}$ and $\delta \sim \sigma^*/\sqrt{N_\bot} \sim 1.4 \times 10^{-10}\mbox{m}$ which satisfies the condition to be much greater than $\lambda_T$. This allows also to estimate the beam density as $n_b \sim \delta^{-3} \sim 3.5\times10^{29} \mbox{m}^{-3}$. In order to evaluate the total number of particles in the beam we use the beam density definition, i.e. $n_b = N/\pi\sigma^{*2}\sigma_z$, obtaining: $N \sim 2.7 \times (10^{9} - 10 ^{10})\,. $ The estimate of the total number $N_z$ of particles along $z$ comes from: $N_z=N/N_\bot \sim 2.2\times (10^6 - 10^7)$. Then the \textit{transverse density} $n_\perp=N_\bot/\pi\sigma^{*2}\sim 1.6\times 10^{19}$ m$^{-2}$ and \textit{longitudinal density} $n_z=N_z/\sigma_z\sim 2.2\times 10^{10}$ m$^{-1}$, respectively.

Supposing that the beam is reaching the lowest QWM state by cooling down from higher temperature, it is instructive to know that how would be the temperature at that state, if we identify $\epsilon_N^*$ as the lowest normalized emittance. Actually, according to our heuristic approach, it would be only an estimate to figure out whether the thermal effects are compatible or not with the visibility of the quantum ones. To this end, let us observe that the condition to impose such an identification is (Maxwellian beam)
\begin{equation}\label{case1}
\frac{\epsilon_N^{*2}}{4\sigma^{*2}}=\frac{k_B\gamma_0T}{m_0c^2}.
\end{equation}
For the above beam energy, emittance and spotsize, Eq. (\ref{case1}) gives the following temperature estimates:
$$T_\perp \sim 5.3\times (10^2 - 10^3)\mbox{K}\,.$$
The above argument justifies our assumption to consider the beam sufficiently \textit{cold} to take into account the individual quantum nature of the particles (single-particle uncertainty principle and spin), but sufficiently \textit{warm} to disregard the collective quantum nature of the beam particles due to overlapping of the wave function (exchange effects).

\section{Conclusions and remarks}
In this paper, we have compared two different wave theories describing the paraxial propagation of a relativistic electron or positron beam in the presence of the collective space-charge effects that take place in a strongly magnetized plasma as well as in a vacuum. This has been done by taking into account the beam propagation in thermal (TWM) as well as quantum (QWM) regimes with a unified quantum formalism. The analysis has been carried out in the strong nonlocal regime that, for sufficiently sharp cylindrically symmetric beams, allows to expand the effective collective potential in powers of the radial coordinate up to the second one. This way, in both theories, the transverse beam dynamics is governed by a 2D Schr\"{o}dinger-like equation. An envelope equation of the Ermakov-Pinney type has been derived for both TWM and QWM regimes. We have shown that in thermal regime it recovers the well known Sacherer equation that governs the time evolution of the beam spot size in the presence of collective effects. In quantum regime, within Hartree's mean field approximation, it describes the evolution of the single-particle spot size, i.e., single quantum ray spot in the transverse plane (Compton regime), in the presence of collective effects, as well. By using heuristic arguments in phase space, involving both the Fermionic nature of the beam particles (electrons or positrons) and their individual quantum nature (quantum uncertainty), we have shown that such evolution contains the same information carried out by an evolution equation for the beam spot size (description of the beam as a whole). Such an evolution equation exhibits, through the collective interaction, an amplification of the single-particle quantum effects. Within this framework we have heuristically described the possibility to reach the lowest quantum state compatible with the Fermionic nature and the negligibility of the single-particle wave function overlapping. On the basis of a set of parameters available in the advanced accelerating machines, we have shown that this quantum collective state is also compatible to the feasibility to reach the nano-sized beams. According to our heuristic description, this lowest QWM state seems to play the role of borderline between the description in fundamental single-particle Compton regime and the collective quantum and thermal regimes at nano- and micro-scales.

Remarkably, nanobeams, whose spot size is roughly ranging from a few $0.1$ to a
few $10$ nm, can allow the quantum particle diffraction to
manifest at macroscopic level. In fact, provided that its
transverse temperature is suitably kept low, the beam dispersion
due to thermal spreading among the particles (thermal emittance)
may be dominated by the one originated by the particle quantum
diffraction. This physical condition may be amplified by the
collective effects that are driven by the beam itself by through
the mechanism of the PWF excitation or collective effects in vacuo.
Nanobeams are currently
employed for a number of scientific and technological
applications, such as microscopy
\cite{Batson2002,Batson2003,Liu2007,Beche2010,Sayed2008},
nanolithography \cite{Zobelli2006}, shaping nanotubes
\cite{Zobelli2007,Zobelli2008}, the study of nonlocal elasticity
of nonlinear mechanical and magnetic vibrations
\cite{Yang2009,Firouz-Abadi2012}, and in the high-energy particle
acceleration
\cite{Biryukov2002,Bellucci2003,Bellucci2005,Assmann2003a,Assmann2003b,
Redaelli2003,Redaelli2004,Saito2003}. In the latter, special
attention has been devoted to the channelling of high-energy beams
in nanotubes \cite{Biryukov2002,Bellucci2003,Bellucci2005} while
recent studies of feasibility of CLIC  (Compact Linear
Colliders) \cite{Zimmermann2006} at CERN propose an
electron-positron linear collider with nanometer-size colliding
beams at an energy of 3 TeV c.m. \cite{Guignard2000}. The CLIC
stability study demonstrated that colliding nanobeams are feasible
with regard to ground motion and stabilization.  The transport,
demagnification and collision of these nanobeams imposes magnet
vibration tolerances that range from 0.2 nm to a few nanometers
which is well below the floor vibration usually observed
\cite{Assmann2003a,Assmann2003b}. Remarkably, the nanometer-sized
transverse spot of the electron or positron beams is also required
to achieve the luminosity at the interaction point of about
$10^{35}$ cm$^{-2}$ s$^{-1}$. An earlier application of the PWF
excitation-based plasma lens theory \cite{Su1990} to CLIC has
preliminarily proven the possibility to achieve very high
luminosity by focusing an electron/positron beam down from 630 to
12 nm in both planes \cite{Fedele1990a}. Furthermore, a laser
wake-field accelerator based on a solid-state plasma was recently
proposed \cite{Saito2003}. In such a device, a large amplitude
plasmon is excited in the inner wall of a metal tube using,
instead of the usual TW laser, a MW laser to produce. Such a
scheme seems to provide a linac to accelerate electron beams with
gradients exceeding the GeV/m level (the so-called
\textit{plasmonlinac}). The electron beam should be provided by a
carbon-nanotube \cite{Saito2000}, so that this device is capable
of producing nanobeams.

\end{document}